# STUDY OF DISTRIBUTION AND ASYMMETRY OF SOLAR ACTIVE PROMINENCES DURING SOLAR CYCLE 23

Navin Chandra Joshi*, Neeraj Singh Bankoti, Seema Pande, Bimal Pande and Kavita Pandey

*Department of Physics, DSB Campus, Kumaun University, Naini Tal – 263 002, Uttarakhand, India*

**E-mail address: njoshi98@gmail.com*

## ABSTRACT

In this paper we present the results of a study of the spatial distribution and asymmetry of solar active prominences (SAP) for the period 1996-2007 (solar cycle 23). For more meaningful statistical analysis we have analysed the distribution and asymmetry of SAP in two subdivisions viz. Group1 (ADF, APR, DSF, CRN, CAP) and Group2 (AFS, ASR, BSD, BSL, DSD, SPY, LPS). The north-south (N-S) latitudinal distribution shows that the SAP events are most prolific in the 21-30° slice in the northern and southern hemispheres and east-west (E-W) longitudinal distribution study shows that the SAP events are most prolific (best visible) in the 81-90° slice in the eastern and western hemispheres. It has been found that the SAP activity during this cycle is low compared to previous solar cycles. The present study indicates that during the rising phase of the cycle the number of SAP events were roughly equal on the north and south hemispheres. However, activity on the southern hemisphere has been dominant since 1999. Our statistical study shows that the N-S asymmetry is more significant then the E-W asymmetry.

## 1. INTRODUCTION

Long term observations of various solar activity phenomena indicate that their occurrence in the northern and southern (as well as eastern and western) hemispheres on the solar disk are not uniform, with more events occurring in one or the other hemisphere during certain period of time. This phenomenon is referred to as asymmetry. The north-south (N-S) and east-west (E-W) distribution and asymmetries, of several solar activity phenomena such as flares, filaments, magnetic flux, relative sunspot numbers, coronal mass ejections (CMEs) and sunspot areas have been investigated by various authors (Maunder, 1904; Howard, 1974; Knoška, 1985; Verma, 1987; Vizoso and Ballester 1987; Verma, 1993; Oliver and Ballester, 1994; Verma, 2000a; Verma, 2000b; Temmer et al., 2001; Joshi and Pant 2005; Gao, Li, and Zhong, 2007). Much work has been done to study the distribution and asymmetry since Maunder (1904), observed and presented the N-S asymmetry of sunspots during the period 1874-1902. Verma (2000b) investigated the N-S and E-W distribution and asymmetries of the solar active prominences (SAP) events for the whole disk for the period 1957-1998 in considerable details. Many authors paid particular attention to the asymmetry of the photospheric features (sunspot relative number, sunspot area, magnetic classes of sunspots etc.) and their relation to the phase of the 11-year solar cycle (SC). Vizoso and Ballester (1987) presented the results of a study of the N-S asymmetry in sudden disappearances of solar prominences (SDP) during solar cycle 18-21. The asymmetries of all solar active features on the entire solar atmosphere were also studied. Verma (1987) studied the N-S asymmetry for major flares, type II radio bursts, white light flares, gamma ray bursts and hard X-ray bursts. Brajša et al. (2005) analysed spatial distribution and N-S asymmetry of coronal bright points from mid 1998 to mid 1999.

Some of the authors (Carbonell, Oliver, and Ballester, 1993; Li, Schmieder, and Li, 1998; Ataç and Özgüç, 2001) demonstrated that the N-S asymmetry has high statistical significance. The E-W asymmetry of solar phenomena also has been studied by various authors (Letfus, 1960; Letfus and Růžičková-Topolvá, 1980; Joshi, 1995) and existence of a small asymmetry has been reported. This means that the non-uniformity of the solar activity (N-S asymmetry in particular) is a real feature and cannot be due to random fluctuations generated from a binomial or uniform distribution of probability between hemispheres. Other authors tried to find a periodicity in this distribution. First, an 11-12 year periodicity was inferred, but whether or not it is related to the classical sunspot cycle is still controversial. Nevertheless most of them calculated that the asymmetry is not in phase with the 11-years SC (Garcia, 1990; Vizoso and Ballester, 1990; Temmer et al., 2001). Long term periods were also suggested; 8 SCs (Vizoso and Ballester, 1990; Ataç and Özgüç, 1996) and even 12 SCs (Verma, 1992; Li et al., 2002). Based on such studies the asymmetry of the solar activity in the SC 23 should favour the southern hemisphere.

In the present paper we investigate the spatial distribution and asymmetry of SAP for the period 1996-2007 (SC 23). In Section 2 we present the observational data and analysis. In Section 3 the latitudinal distributions and N-S asymmetry are discussed. In Section 4 the longitudinal distributions and E-W asymmetry are discussed. Our approach consists of examining the ascending, maximum and descending phase of SC 23. In Section 5 we have presented the comparison between SC 20, 21, 22, and 23 and in the final section (6) results and discussions have been presented.

## 2. OBSERVATION DATA AND ANALYSIS

The data used in the present study have been collected from National Geophysical Data Center's (NGDC's) anonymous ftp server during 01 January 1996 to 31 December 2007. This period covers SC 23. The URL address of this website is as follows: *ftp://ftp.ngdc.noaa.gov/STP/SOLAR_DATA/SOLAR_FILAMENTS.* The SAP data include limb and disk features and events. During this period of 4383 days, the occurrence of a total number 8778 of SAP is reported. From this database those events have been excluded which occurred at 0º latitude and central meridian distance (CMD). After excluding such events, we get a total of 873 2 SAP for N-S distribution and 8712 SAP for E-W distribution. Table 1 lists different limb and disk features and their corresponding percentage.

Table 1. The number of different limb and disk features and the corresponding percentage values during SC 23.

| Limb and Disk features | | Group Number | Number of events | % |
|---|---|---|---|---|
| DSF | (Disappearing filament) | 1 | 1962 | 22.35 |
| AFS | (Arch filament system) | 2 | 1923 | 21.91 |
| DSD | (Dark surge on disc) | 2 | 1488 | 16.95 |
| ADF | (Active dark filament) | 1 | 1202 | 13.69 |
| BSL | (Bright surge on limb) | 2 | 600 | 6.84 |
| ASR | (Active surge region) | 2 | 574 | 6.54 |
| APR | (Active prominence) | 1 | 344 | 3.92 |
| EPL | (Eruptive prominence on limb) | | 336 | 3.83 |
| LPS | (Loops) | 2 | 155 | 1.77 |
| BSD | (Bright surge on disc) | 2 | 146 | 1.66 |
| SPY | (Spray) | 2 | 33 | 0.38 |
| CAP | (Cap prominence) | 1 | 14 | 0.16 |
| CRN | (Coronal rain) | 1 | 1 | 0.01 |
| SSB | (Solar sector boundary) | | 0 | 0.00 |
| MDP | (Mound prominence) | | 0 | 0.00 |
| Total | | | 8778 | 100.00 |

From this table we can see that the DSF is the most dominating feature whereas DSF, AFS, DSD and ADF together give 74.90% of total SAP events during the period of study of SC 23. For the more meaningful statistical analysis we have analysed different features separately along with the total SAP data. For this we have formed two groups related as far as physics is concerned: Group 1 (G1) for structures belonging to prominence/filament and Group 2 (G2) to active regions. The EPL data have been left out because EPL is an ambiguous group which contains eruptions of prominences as well as solar flare eruptions close to the limb and can not be distributed equally amongst the two subgroups chosen. Table 2 lists the number of features in two subgroups and their percentage values. It shows that G2 consists of more number of features (58.27%) compared to G1 (41.73%). The yearly variation of different features during the SC 23 can be clearly seen from Table 7.

Table 2. Number of events in the two subgroups and the corresponding percentage values during SC 23.

| Group | Number | % |
|---|---|---|
| G1: ADF, APR, DSF, CRN, CAP | 3523 | 41.73 |
| G2: AFS, ASR, BSD, BSL, DSD, SPY, LPS | 4919 | 58.27 |
| Total | 8442 | 100.00 |

In Figure 1, the monthly number of SAP, monthly number of G1 and G2 and monthly mean sunspot number (SN) during the period of our investigation have been plotted. This figure also represents the plots for monthly number of Hα solar flares and subflares (SF) as well as for solar flare index (Q) therein. We have included all Hα flare events having importance equal and greater than S (subflare). The flare index (Q) represents daily flare activity observed during 24 hours of the day. It is calculated for each flare using the formula $Q = i \times t$, where $i$ is the importance coefficient of the flare and $t$ is the duration of the flare in minutes (Kleczek, 1952). SF, SN data and the calculated values of Q are available in anonymous ftp servers of NGDC. The URL address of these websites is as follows: *ftp://ftp.ngdc.noaa.gov/STP/SOLAR_DATA/SOLAR_FLARES/INDEX* and *ftp://ftp.ngdc.noaa.gov/STP/SOLAR_DATA/SUNSPOT_NUMBERS.*

From Figure 1 we can see that the variation of Q, SF and SN is similar but different from the SAP as well as for G1 and G2. During the ascending phase of SC 23, SAP, G1 and G2 are maximum in number, but other phenomena i.e. SF and SN are minimum in number. However, after 1999 this contrasting appearance (in SAP and SF, SN) is not exhibited. During the year 1996 and beyond 1998 G1 shows similar variation as SAP, while G2 shows a similar behavior all along as that of SAP barring the peak height before 1997. Figure 1 shows peaks during the maximum phase (2000-2001) of the SC under investigation.

## 2.1 STATISTICAL ANALYSIS

To specify the statistical significance of the N-S and E-W asymmetry indices, we followed Letfus (1960) and Letfus and Růžičkoυá-Topoloυá (1980). We have calculated the N-S asymmetry indices $(A_{NS})$ and E-W asymmetry indices $(A_{EW})$ by using the formula

$$A_{NS} = \frac{N-S}{N+S}, \; A_{EW} = \frac{E-W}{E+W} \qquad (1)$$

The dispersion of the N-S and E-W asymmetry of a random distribution of flares is given by

$$\Delta A_{NS} = \pm \frac{1}{\sqrt{2(N+S)}}, \qquad (2)$$

$$\Delta A_{EW} = \pm \frac{1}{\sqrt{2(E+W)}} \qquad (3)$$

Here $N$ and $S$ are the numbers of SAP observed in the northern and southern halves of the solar disk and $E$, $W$ are the number of SAP observed in the eastern and western halves respectively. Thus, if $A_{NS} > 0$ the activity in the northern hemisphere dominates or else it will dominate the southern hemisphere, and if $A_{EW} > 0$ the activity in the eastern hemisphere dominates otherwise it will dominate the western hemisphere.

To verify the reliability of the observed N-S and E-W asymmetry values, a $\chi^2$ - test is applied with Equations (4) and (5) given below respectively.

$$\chi_{NS} = \frac{2(N+S)}{\sqrt{(N+S)}} = \frac{\sqrt{2}A_{NS}}{\Delta A_{NS}}, \qquad (4)$$

$$\chi_{EW} = \frac{2(E+W)}{\sqrt{(E+W)}} = \frac{\sqrt{2}A_{EW}}{\Delta A_{EW}} \qquad (5)$$

If $A_{NS} = \Delta A_{NS}$ and $A_{EW} = \Delta A_{EW}$ the probability that the asymmetry exceeds the dispersion value is p = 84%, and if $A_{NS} = 2\Delta A_{NS}$ and $A_{EW} = 2\Delta A_{EW}$ p is 99.5 %. Here the first limit differentiates between the statistically significant and insignificant values and the second separates the values which may be considered quite realistic. The limits divide the values of the asymmetry into three categories: with low, intermediate and high probabilities (Joshi, 1995; Temmer et al., 2001; Joshi and Joshi, 2004).

## 3. LATITUDINAL DISTRIBUTION AND N-S ASYMMETRY

The SAP data obtained from NGDC were analyzed to understand the N-S distribution and N-S asymmetry. In Table 3 we have shown the yearly number of SAP events at an interval of 10° in the northern and southern hemispheres. Column 12 of Table 3 gives the total number of SAP in northern and southern hemispheres. In Figure 2(a), we have plotted the number of SAP, G1 and G2 versus heliographic latitude in degree for SC 23. The 0° latitude represents the equator of the Sun. Form Figure 2(a) it is clear that the SAP as well as G1 and G2 activity is maximum between 21-30° latitude in each hemisphere. The N-S asymmetry indices for SAP as also for G1 and G2 based on annual counts from 1996 to 2007 have been plotted in Figure 2(b). From this figure it is clear that all curves show similar behavior. In 1997 the cycle was northern hemisphere dominated but after 1999 it becomes southern dominated and stayed there for the remaining years. 8 out of 12 N-S asymmetry values turn out to be highly significant with a probability $p \geq 99.5\%$ where the observed asymmetry index exceeds the dispersion value of a random distribution. 1 and 3 out of 12 values come out to be statistically significant and insignificant respectively. The highly significant, significant and insignificant values of N-S asymmetry indices are marked with black squares, black stars and white circles respectively in Figure 2(b).

## 4. LONGITUDINAL DISTRIBUTION AND E-W ASYMMETRY

The data downloaded from NGDC have also been used to study the E-W distribution of SAP as well as G1 and G2 data for the period 1996-2007. In Table 4 we have shown the yearly distribution of SAP events at CMD intervals of 10° from the central meridian towards the east and west limbs respectively during cycle 23. To understand the table we have plotted the number of SAP, G1 and G2 versus heliographic CMD in degrees in Figure 3(a). In Figure 3(a) the minus (-) sign in heliographic CMD indicates eastward and plus (+) sign represents westward, here -90° represent E90 and +90° represent W90. From Figure 3(a), it is clear that the number of SAP as well as G1 and G2 events decreases from 1° to 80° and then SAP, G1 and G2 frequency increases between 81-90° CMD. The E-W asymmetry indices for all SAP and the two subgroups i.e., G1 and G2 versus year graph have been plotted in Figure 3(b). From this figure it can be seen that all curves show same type of behavior. 5 out of 12 asymmetry values reveal a high statistical significance with $p \geq 99.5\%$. 4 and 3 out of 12 come out to be significant and statistically insignificant respectively. The highly significant, significant and insignificant values of E-W asymmetry indices are marked with black squares, black stars and white circles respectively in Figure 3(b).

## 5. COMPARISION AMONG SOLAR CYCLE 20, 21, 22 AND 23

Verma (2000b) has calculated the yearly number of the SAP events in the interval of 10° latitude for northern and southern hemispheres as well as for the eastern and western hemispheres respectively from 1957 to 1998. In Tables 5 and 6 of our paper, we have counted and presented the total number of SAP events for SCs 20, 21, 22 and 23 in the interval of 10° latitude and CMD from 0° to 90°. We have considered the data from 1963 to 2007 covering four SCs i.e. 20, 21, 22 and 23. The N-S and E-W asymmetry of SAP events is shown by the solid lines in Figure 4 and 5 with highly significant, significant and insignificant values, marked with black squares, black stars and white circles respectively. 37 out of 45 N-S asymmetry values reveal a high statistical significance with $p \geq 99.5\%$. 1 and 7 come out to be statistically significant and insignificant respectively. 29 out of 45 E-W asymmetry values reveal a high statistical significance with $p \geq 99.5\%$ whereas 10 and 6 out of 45 turn out to be significant and statistically insignificant respectively. In Figure 4, we have also plotted the N-S asymmetry in the SDP for the period 1947-1985 (Vizoso and Ballester, 1987) by the dotted line.

## 6. DISCUSSIONS AND CONCLUSIONS

In the present study the distribution of SAP, G1 and G2 in the northern and southern (as well as eastern and western) hemispheres for SC 23 have been analyzed and the results obtained are as follows:

- From the equator of the Sun, frequency of SAP, G1 and G2 events increases from 1° to 30° and then decreases from 31° to 90° and the SAP events are maximum between latitudes 21-30° for SC 23 but for the SCs 20, 21 and 22 SAP activity was maximum between 11-20° latitude band on either side of the solar equator.
- From the central meridian of the Sun, the frequency of SAP, G1 and G2 events decreases from 1° to 80° and frequency increases between 81° to 90°. The SAP events are most numerous between 81° to 90° CMD band.
- From the Table 5 and 6 it has been found that the SAP activity during this cycle is low compared to previous four SCs.
- From the statistical analysis it is clear that, N-S asymmetry is more significant than the E-W asymmetry.

On comparing latitudinal distribution during SC 23 (Figure 2(a)) with four previous SCs (see Figure 3 in Verma, 2000b) it can be seen that SAP events are most prolific in 21-30° different from the previous SCs (11-20°), whereas the CMD distribution (Figure 3(a)) shows similar behavior (see Figure 1 in Verma, 2000b). For the SC 23 the SAP events are not uniformly distributed in northern and southern hemispheres. The southern dominance of solar activity during SC 23 has been investigated by several authors by taking different solar activity features i.e. solar flares and CMEs (Joshi, Pant, and Manoharan, 2006; Gao, Li, and Zhong, 2007).

Similar trend has been found in the present investigation with SAP data. The present study and the study of Ataç and Özgüç (1996) confirm the predictions of Verma (1992) for SC 23. According to Verma (1992) the N-S asymmetry of solar active phenomena may be southern dominated during SCs 22, 23, and 24 and will be northern dominated during SC 25. Temmer et al. (2001) presented a statistical analysis of Hα flare from 1975 to 1999 and found that there exists a significant N-S asymmetry and slight but significant E-W asymmetry. Similar result has been found in our study also.

Tables 3 and 4 show many interesting aspects of SAP distribution (latitude and CMD) with the evolution of SC 23. From Table 3 it is clear that in the ascending phase of the cycle the number of SAP first increases up to 1997 and then decreases from 1998 to the end of the cycle under investigation. In the beginning of the cycle 1-10° latitude produced maximum number of SAP events. In the succeeding years after 1996, most of the SAP were located between 21-30° latitude belt and with the progress of the cycle the number of SAP decreased in lower altitudes. Table 4 shows that in the beginning of the cycle the frequency of SAP was most prolific both in low (1-10°) as well as in high (81-90°) CMD bands. In the year 1997, just after the solar minimum, most of SAP events were located between 21-30° and 81-90° CMD bands. It is clear from Tables 5 and 6 that indices of N-S and E-W asymmetry of SAP events favour the northern and western hemisphere for the SCs 20 and 21 and shift to southern and eastern hemisphere during cycles 22 and 23. Table 7 represents the yearly variation of different features of SAP during SC 23. It is clear from this table that most of the SAP events (69.08%) occur during the rising phase (1996, 1997 and 1998) of the cycle under investigation. From all SAP events AFS, DSD, ADF, BSL, ASR, APR and BSD are maximum in number during the rising phase and decrease as the cycle progresses whereas DSF, EPL, LPS and SPY are minimum during the rising phase and become maximum during the maximum phase. On comparing our results with Verma (2000b) we find the variation of N-S asymmetry index during cycle 23 differs from previous three cycles i.e. 20, 21 and 22. From Figure 4 and 5 it is clear that the N-S and E-W asymmetry do not show any systematic behavior and have no relation with CSs maxima and minima during 1963 to 2007.

**ACKNOWLEDGEMENTS**

Authors thank UGC, New Delhi for financial assistance under DSA (phase-III) program running in the Department of Physics, Kumaun University, Nainital. Two of the authors NCJ and NSB are thankful to UGC, New Delhi for financial assistance under RFSMS (Research Fellowship in Science for meritorious students) scheme. The authors are also thankful to the referee and editor for their valuable comments and suggestions.

## REFERENCES

Ataç, T., and Özgüç, A.: 1996, Solar Phys. **166**, 201.

Ataç, T., and Özgüç, A.: 2001, Solar Phys. **198**, 399.

Brajša, R., Wöhl, H., Vršnak, B., Ruždjak, V., Clette, F., Hochedez, J.-F., Verbanac, G., Temmer, M.: 2005, Solar Phys. **231**, 29.

Carbonell, M., Oliver, R., and Ballester, J. L.: 1993, Astron. Astrophys. **274**, 497.

Garcia, H.A.: 1990, Solar Phys. **127**, 185.

Gao, P. X., Li, Q. X., and Zhong, S. H.: 2007, J. Astrophys. Astr. **28**, 207.

Howard, R.: 1974, Solar Phys. **38**, 59.

Joshi, A.: 1995, Solar Phys. **157**, 315.

Joshi, B., and Joshi, A.: 2004, Solar Phys. **219**, 343.

Joshi, B. and Pant, P.: 2005, Astron. Astrophys. **431**, 359.

Joshi, B., Pant, P., and Manoharan, P. K.: 2006, J. Astrophys. Astr. **27**, 151.

Kleczek, J.: 1952, Publ. Czech. Centr. Astron. Inst., No.22.

Knoška, Š.: 1985, Contrib. Astron. Obs. Skalnaté pleso **13**, 217.

Letfus, V.: 1960, Bull. Astron. Inst. Czech. **11**, 31.

Letfus, V., Růžičkoυá-Topoloυá, B.: 1980, Bull. Astron. Inst. Czech. **31**, 232.

Li, K. -J., Schmieder, B., and Li, Q. -Sh.: 1998, Astron. Astrophys. Suppl. Ser. **131**, 99.

Li, J. K., Wang, J. X., Xiong, S. Y., Liang, H. F., Yun, H. S., and Gu, X. M.: 2002, Astron. Astrophys. **383**, 648.

Maunder, E.W.: 1904, Monthly Notices Roy. Astron. Soc. **64**, 747.

Oliver, R., and Ballester, J. L.: 1994, Solar Phys. **152**, 481.

Temmer, M., Veronig, A., Hanslmeier, A., Otruba, W., Messerotti, M.: 2001, Astron. Astrophys. **375**, 1049.

Verma, V. K.: 1987, Solar Phys. **114**, 185.

Verma, V. K.:1992, Astron. Soc. Pacific Conf. Ser. **27**, 429.

Verma, V. K.: 1993, Astrophys. J. **403**, 797.

Verma, V. K.: 2000a, J. Astrophys. Astr. **21**, 173.

Verma, V. K.: 2000b, Solar Phys. **194**, 87.

Vizoso, G., and Ballester, J. L.: 1987, Solar Phys. **112,** 317.

Vizoso, G., and Ballester, J. L.: 1990, Astron. Astrophys. **229**, 540.

---------------

**Table 3.** Number of SAP at different latitude bands in the northern (N) and southern (S) hemispheres and tabulated for each year. The dominant hemisphere (DH) and asymmetry index (A-Index) are given for all the years as well as for all the latitudinal bands. SAP that occurred at the equator have been excluded.

| Years | | Number of S AP | | | | | | | | | Total | A-Index | DH |
|---|---|---|---|---|---|---|---|---|---|---|---|---|---|
| | | 1-10° | 11-20° | 21-30° | 31-40° | 41-50° | 51-60° | 61-70° | 71-80° | 81-90° | | | |
| 1996 | N | 611 | 194 | 91 | 41 | 31 | 15 | 11 | 8 | 12 | 1014 | -0.08360 | S |
| | S | 653 | 323 | 110 | 40 | 25 | 11 | 9 | 16 | 12 | 1199 | | |
| 1997 | N | 194 | 398 | 657 | 173 | 27 | 6 | 2 | 0 | 4 | 1461 | +0.16787 | N |
| | S | 92 | 178 | 581 | 151 | 32 | 3 | 0 | 3 | 1 | 1041 | | |
| 1998 | N | 12 | 157 | 231 | 44 | 11 | 3 | 1 | 0 | 1 | 460 | -0.30250 | S |
| | S | 12 | 230 | 427 | 119 | 50 | 17 | 0 | 2 | 2 | 859 | | |
| 1999 | N | 15 | 61 | 72 | 40 | 19 | 8 | 6 | 1 | 1 | 223 | +0.00677 | N |
| | S | 19 | 72 | 55 | 35 | 19 | 14 | 6 | 0 | 0 | 220 | | |
| 2000 | N | 39 | 94 | 95 | 25 | 23 | 9 | 1 | 0 | 1 | 287 | -0.02547 | S |
| | S | 48 | 106 | 78 | 39 | 20 | 10 | 1 | 0 | 0 | 302 | | |
| 2001 | N | 35 | 64 | 63 | 37 | 13 | 1 | 0 | 0 | 0 | 213 | -0.10879 | S |
| | S | 50 | 81 | 70 | 35 | 20 | 7 | 0 | 1 | 1 | 265 | | |
| 2002 | N | 41 | 61 | 34 | 22 | 11 | 4 | 0 | 0 | 0 | 173 | -0.15610 | S |
| | S | 38 | 86 | 65 | 28 | 14 | 5 | 0 | 1 | 0 | 237 | | |
| 2003 | N | 55 | 37 | 28 | 28 | 9 | 2 | 1 | 0 | 0 | 160 | -0.13747 | S |
| | S | 64 | 79 | 39 | 18 | 6 | 3 | 0 | 0 | 2 | 211 | | |
| 2004 | N | 30 | 22 | 13 | 4 | 0 | 1 | 0 | 0 | 0 | 70 | -0.14634 | S |
| | S | 38 | 36 | 10 | 4 | 1 | 3 | 0 | 0 | 2 | 94 | | |
| 2005 | N | 20 | 28 | 5 | 3 | 3 | 0 | 0 | 0 | 2 | 61 | -0.04688 | S |
| | S | 20 | 26 | 8 | 2 | 3 | 2 | 3 | 0 | 3 | 67 | | |
| 2006 | N | 11 | 5 | 8 | 1 | 3 | 2 | 0 | 0 | 1 | 31 | -0.39216 | S |
| | S | 17 | 16 | 14 | 17 | 4 | 2 | 0 | 0 | 1 | 71 | | |
| 2007 | N | 3 | 2 | 0 | 0 | 0 | 0 | 0 | 0 | 0 | 5 | -0.23077 | S |
| | S | 5 | 0 | 1 | 0 | 0 | 0 | 0 | 0 | 2 | 8 | | |
| Total | N | 1066 | 1123 | 1297 | 418 | 150 | 51 | 22 | 9 | 22 | 4158 | -0.04764 | S |
| | S | 1056 | 1233 | 1458 | 488 | 194 | 77 | 19 | 23 | 26 | 4574 | | |
| A-Index | | +0.0047 | -0.0467 | -0.0584 | -0.0772 | -0.1280 | -0.2031 | +0.0732 | -0.4375 | -0.0833 | -0.0476 | | |
| DH | | N | S | S | S | S | S | N | S | S | S | | |

**Table 4.** Number of SAP at different CMD bands in the eastern (E) and western (W) hemispheres and tabulated for each year. The dominant hemisphere (DH) and asymmetry index (A-Index) are given for all the years as well as for all the CMD bands. SAP that occurred at the equator have been excluded.

| **Years** | | **Number of SAP** | | | | | | | | | **Total** | **A-Index** | **DH** |
|---|---|---|---|---|---|---|---|---|---|---|---|---|---|
| | | **1-10°** | **11-20°** | **21-30°** | **31-40°** | **41-50°** | **51-60°** | **61-70°** | **71-80°** | **81-90°** | | | |
| **1996** | **E** | 172 | 133 | 140 | 132 | 94 | 56 | 48 | 26 | 178 | **979** | **-0.11643** | **W** |
| | **W** | 206 | 170 | 163 | 133 | 119 | 86 | 75 | 28 | 257 | **1237** | | |
| **1997** | **E** | 189 | 195 | 207 | 162 | 148 | 125 | 84 | 40 | 140 | **1290** | **+0.03907** | **E** |
| | **W** | 157 | 180 | 177 | 152 | 129 | 108 | 54 | 35 | 201 | **1193** | | |
| **1998** | **E** | 93 | 90 | 74 | 67 | 68 | 50 | 40 | 13 | 120 | **615** | **-0.06606** | **W** |
| | **W** | 73 | 102 | 76 | 65 | 72 | 53 | 34 | 21 | 206 | **702** | | |
| **1999** | **E** | 22 | 21 | 25 | 24 | 16 | 11 | 12 | 11 | 83 | **225** | **+0.01351** | **E** |
| | **W** | 19 | 30 | 17 | 17 | 15 | 19 | 12 | 4 | 86 | **219** | | |
| **2000** | **E** | 33 | 45 | 27 | 30 | 18 | 17 | 8 | 6 | 90 | **274** | **-0.06644** | **W** |
| | **W** | 42 | 34 | 21 | 39 | 23 | 31 | 5 | 7 | 111 | **313** | | |
| **2001** | **E** | 38 | 50 | 19 | 34 | 18 | 10 | 11 | 4 | 66 | **250** | **+0.04384** | **E** |
| | **W** | 39 | 26 | 30 | 31 | 18 | 9 | 6 | 3 | 67 | **229** | | |
| **2002** | **E** | 17 | 15 | 25 | 17 | 19 | 13 | 7 | 2 | 59 | **174** | **-0.14914** | **W** |
| | **W** | 28 | 25 | 18 | 30 | 20 | 9 | 5 | 3 | 97 | **235** | | |
| **2003** | **E** | 24 | 18 | 20 | 19 | 14 | 17 | 13 | 2 | 50 | **177** | **-0.04065** | **W** |
| | **W** | 24 | 32 | 21 | 18 | 21 | 15 | 10 | 2 | 49 | **192** | | |
| **2004** | **E** | 7 | 9 | 16 | 18 | 5 | 3 | 3 | 2 | 18 | **81** | **-0.01220** | **W** |
| | **W** | 10 | 8 | 11 | 6 | 11 | 6 | 2 | 1 | 28 | **83** | | |
| **2005** | **E** | 7 | 8 | 6 | 5 | 7 | 2 | 4 | 4 | 17 | **60** | **-0.07692** | **W** |
| | **W** | 8 | 14 | 6 | 7 | 0 | 5 | 2 | 2 | 26 | **70** | | |
| **2006** | **E** | 11 | 8 | 3 | 5 | 5 | 7 | 1 | 1 | 12 | **53** | **+0.03922** | **E** |
| | **W** | 9 | 9 | 11 | 8 | 1 | 1 | 1 | 0 | 9 | **49** | | |
| **2007** | **E** | 1 | 1 | 0 | 0 | 0 | 0 | 0 | 0 | 2 | **4** | **-0.33333** | **W** |
| | **W** | 3 | 2 | 0 | 0 | 1 | 1 | 1 | 0 | 0 | **8** | | |
| **Total** | **E** | **614** | **593** | **562** | **513** | **412** | **311** | **231** | **111** | **835** | **4182** | **-0.03994** | **W** |
| | **W** | **618** | **632** | **551** | **506** | **430** | **343** | **207** | **106** | **1137** | **4530** | | |
| **A-Index** | | **-0.0033** | **-0.0318** | **+0.0099** | **+0.0069** | **-0.0214** | **-0.0490** | **+0.0548** | **+0.0230** | **-0.1531** | **-0.03994** | | |
| **DH** | | **W** | **W** | **E** | **E** | **W** | **W** | **E** | **E** | **W** | **W** | | |

**Table 5.** Total number of SAP at different latitude bands in the northern (N) and southern (S) hemispheres and tabulated for four SCs. The dominant hemisphere (DH) and asymmetry index (A-Index) are given for four solar cycles. SAP that occurred exactly at the equator have been excluded.

| Cycle | | Total number of SAP | | | | | | | | | Total | A-Index | DH |
|---|---|---|---|---|---|---|---|---|---|---|---|---|---|
| | | 1-10° | 11-20° | 21-30° | 31-40° | 41-50° | 51-60° | 61-70° | 71-80° | 81-90° | | | |
| **20** | **N** | 6039 | 10922 | 7533 | 2285 | 713 | 325 | 220 | 145 | 113 | **28295** | **+0.22487** | **N** |
| | **S** | 5621 | 7175 | 3355 | 886 | 383 | 251 | 98 | 69 | 68 | **17906** | | |
| **21** | **N** | 2553 | 3141 | 1585 | 759 | 350 | 210 | 247 | 281 | 297 | **9423** | **+0.00943** | **N** |
| | **S** | 2343 | 3193 | 1757 | 648 | 351 | 228 | 214 | 259 | 254 | **9247** | | |
| **22** | **N** | 10400 | 12505 | 7064 | 2646 | 760 | 370 | 333 | 311 | 373 | **34762** | **-0.05233** | **S** |
| | **S** | 9246 | 16034 | 8555 | 2875 | 798 | 347 | 240 | 237 | 269 | **38601** | | |
| **23** | **N** | 1066 | 1123 | 1297 | 418 | 150 | 51 | 22 | 9 | 22 | **4158** | **-0.04764** | **S** |
| | **S** | 1056 | 1233 | 1458 | 488 | 195 | 76 | 19 | 23 | 26 | **4574** | | |

**Table 6.** Number of SAP at different CMD bands in the eastern (E) and western (W) hemispheres and tabulated for four SCs. The dominant hemisphere (DH) and asymmetry index (A-Index) are given for four solar cycles. SAP that occurred exactly at the equator have been excluded.

| Cycle | | Total number of SAP | | | | | | | | | Total | A-Index | DH |
|---|---|---|---|---|---|---|---|---|---|---|---|---|---|
| | | 1-10° | 11-20° | 21-30° | 31-40° | 41-50° | 51-60° | 61-70° | 71-80° | 81-90° | | | |
| **20** | **E** | 1948 | 1807 | 1705 | 1487 | 1222 | 863 | 593 | 502 | 13191 | **23318** | **+0.00955** | **E** |
| | **W** | 1839 | 1640 | 1654 | 1409 | 1187 | 815 | 566 | 510 | 13257 | **22877** | | |
| **21** | **E** | 558 | 590 | 496 | 480 | 355 | 318 | 196 | 91 | 6535 | **9619** | **+0.03037** | **E** |
| | **W** | 572 | 526 | 530 | 515 | 377 | 311 | 190 | 103 | 5928 | **9052** | | |
| **22** | **E** | 4453 | 4452 | 4289 | 4078 | 3783 | 3241 | 2546 | 1647 | 10049 | **38538** | **-0.03966** | **W** |
| | **W** | 5074 | 4804 | 4612 | 4307 | 3873 | 3471 | 2562 | 1820 | 11198 | **41721** | | |
| **23** | **E** | 614 | 593 | 562 | 513 | 412 | 311 | 231 | 111 | 835 | **4182** | **-0.03994** | **W** |
| | **W** | 618 | 632 | 551 | 506 | 430 | 343 | 207 | 106 | 1137 | **4530** | | |

**Table 7.** Number of SAP (limb and disk features) tabulated for each year.

| Years | Number of SAP (Limb and Disk features) | | | | | | | | | | | | | | Total |
|---|---|---|---|---|---|---|---|---|---|---|---|---|---|---|---|
| | DSF | AFS | DSD | ADF | BSL | ASR | APR | EPL | LPS | BSD | SPY | CAP | CRN | SSB | MDP | |
| **1996** | 107 | 635 | 528 | 490 | 184 | 194 | 61 | 9 | 2 | 27 | 0 | 1 | 0 | 0 | 0 | **2238** |
| **1997** | 112 | 931 | 572 | 464 | 13 | 258 | 67 | 12 | 2 | 74 | 1 | 0 | 0 | 0 | 0 | **2506** |
| **1998** | 237 | 353 | 242 | 117 | 91 | 93 | 34 | 76 | 29 | 41 | 2 | 5 | 0 | 0 | 0 | **1320** |
| **1999** | 204 | 1 | 31 | 30 | 76 | 10 | 35 | 42 | 10 | 0 | 4 | 3 | 0 | 0 | 0 | **446** |
| **2000** | 318 | 0 | 27 | 37 | 72 | 9 | 52 | 39 | 28 | 0 | 8 | 2 | 1 | 0 | 0 | **593** |
| **2001** | 296 | 0 | 19 | 18 | 41 | 6 | 39 | 24 | 28 | 3 | 5 | 0 | 0 | 0 | 0 | **479** |
| **2002** | 207 | 0 | 23 | 24 | 58 | 3 | 37 | 45 | 11 | 0 | 1 | 3 | 0 | 0 | 0 | **412** |
| **2003** | 247 | 0 | 18 | 8 | 35 | 0 | 4 | 34 | 22 | 0 | 4 | 0 | 0 | 0 | 0 | **372** |
| **2004** | 92 | 1 | 16 | 6 | 15 | 0 | 3 | 20 | 6 | 0 | 6 | 0 | 0 | 0 | 0 | **165** |
| **2005** | 65 | 0 | 7 | 4 | 10 | 1 | 6 | 22 | 15 | 1 | 1 | 0 | 0 | 0 | 0 | **132** |
| **2006** | 70 | 0 | 5 | 4 | 5 | 0 | 6 | 11 | 0 | 0 | 1 | 0 | 0 | 0 | 0 | **102** |
| **2007** | 7 | 2 | 0 | 0 | 0 | 0 | 0 | 2 | 2 | 0 | 0 | 0 | 0 | 0 | 0 | **13** |
| **Total** | **1962** | **1923** | **1488** | **1202** | **600** | **574** | **344** | **336** | **155** | **146** | **33** | **14** | **1** | **0** | **0** | **8778** |

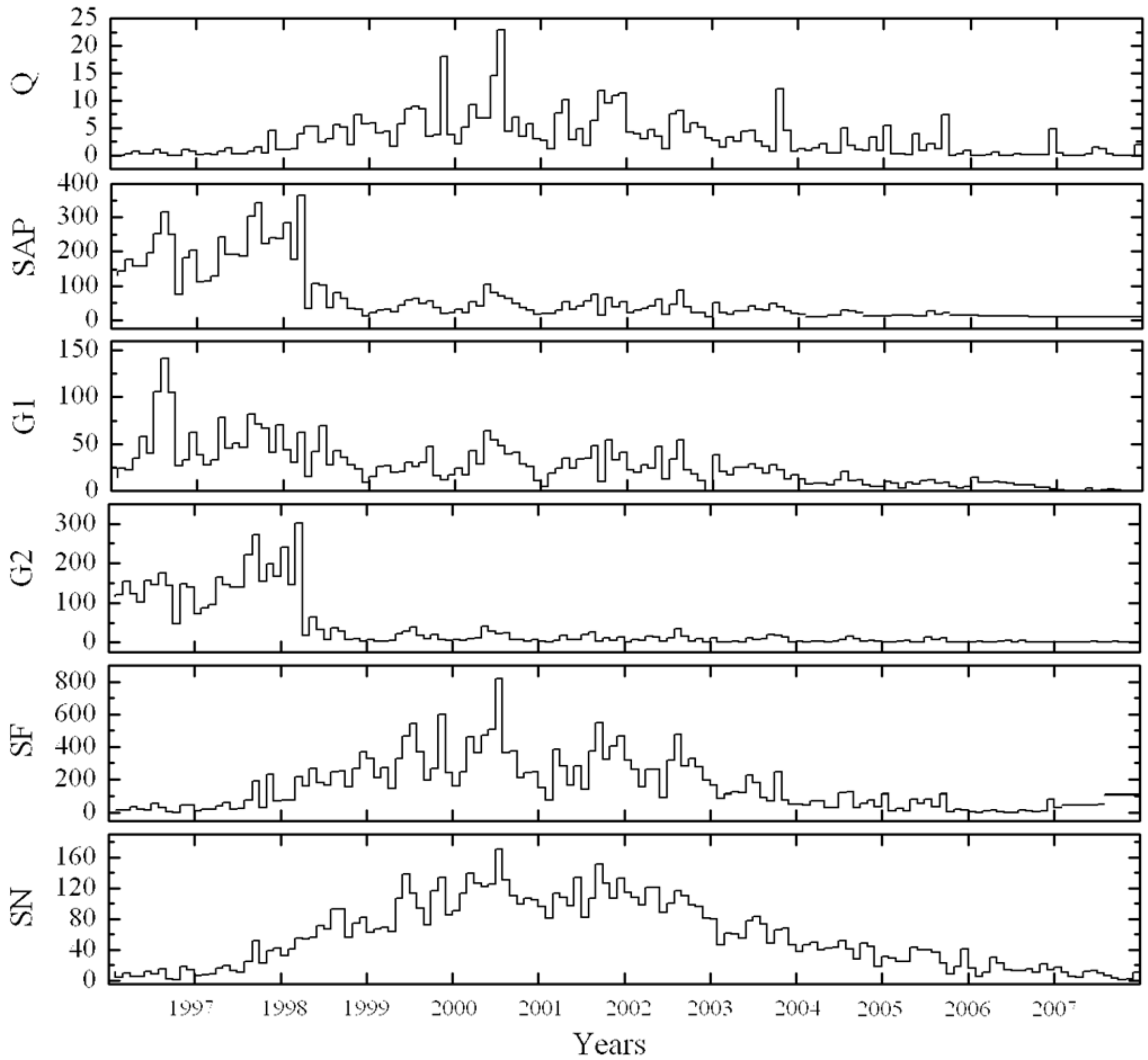

**Figure 1.** Monthly plots of different solar activity parameters, flare index (Q), solar active prominences (SAP), Group1 (G1), Group2 (G2), Hα solar flares and sub-flares (SF) and sunspot number (SN) from 1996-2007 (from top to bottom panel).

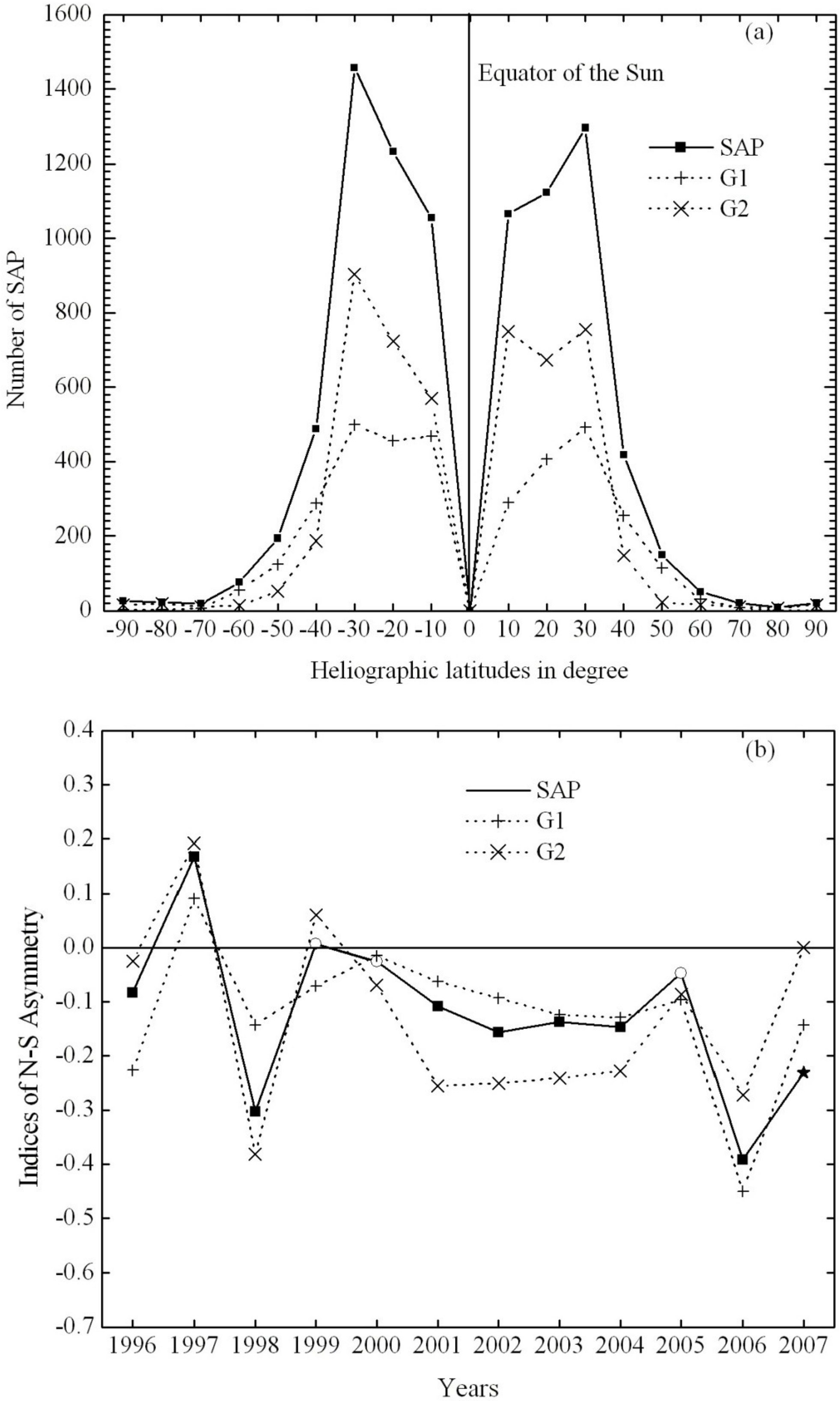

**Figure 2. (a).** Plot of SAP (solid line), G1 and G2 (doted line) versus north-south heliographic latitudes in degrees. (b). Plot of N-S asymmetry indices for SAP (solid line), G1 and G2 (doted line) events versus years (1996-2007). Highly significant, significant and insignificant values are marked with black squares (■), black stars (★) and white circles (○) respectively.

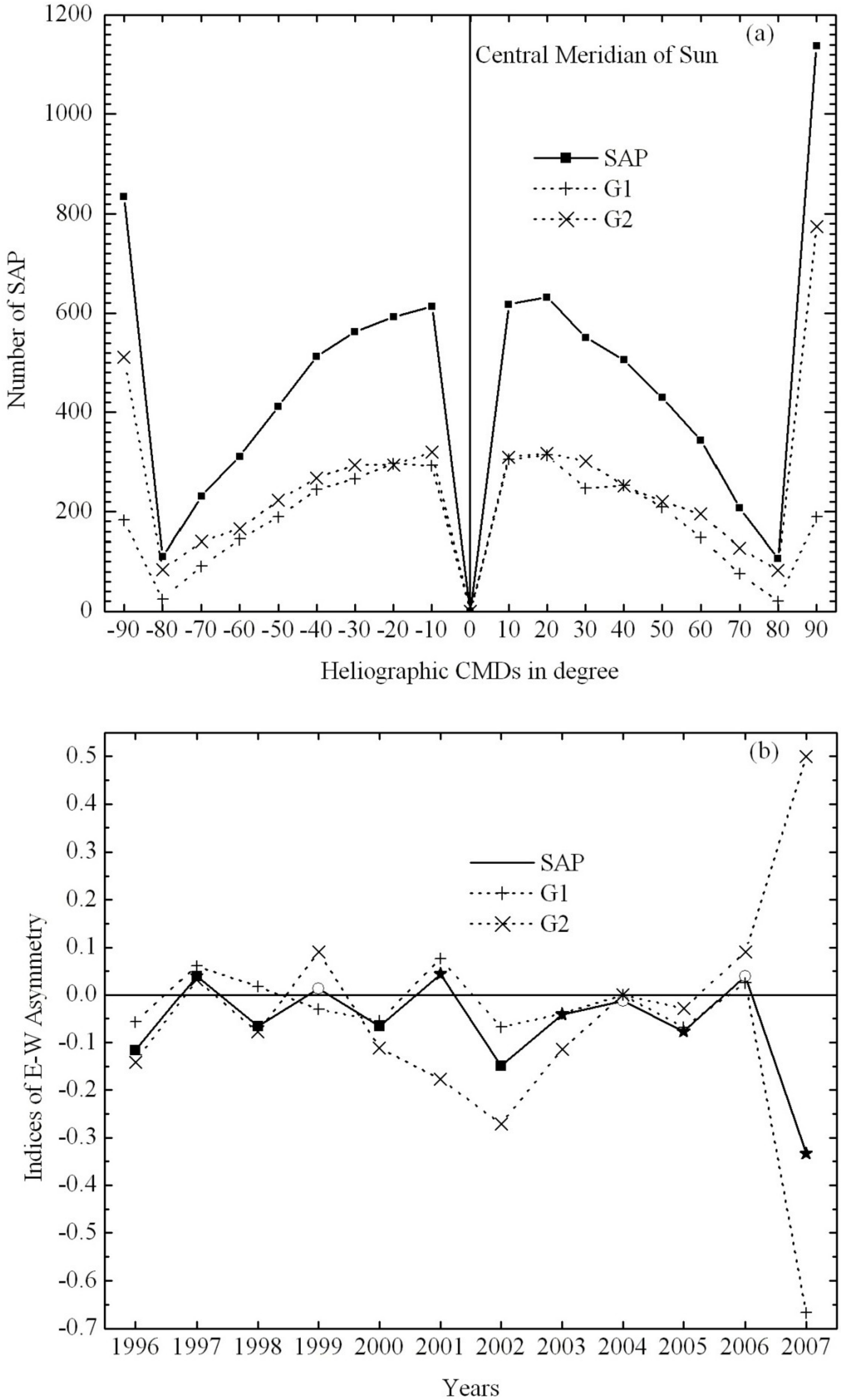

**Figure 3.** Same as Figure 2, but for CMDs distribution and E-W asymmetry.

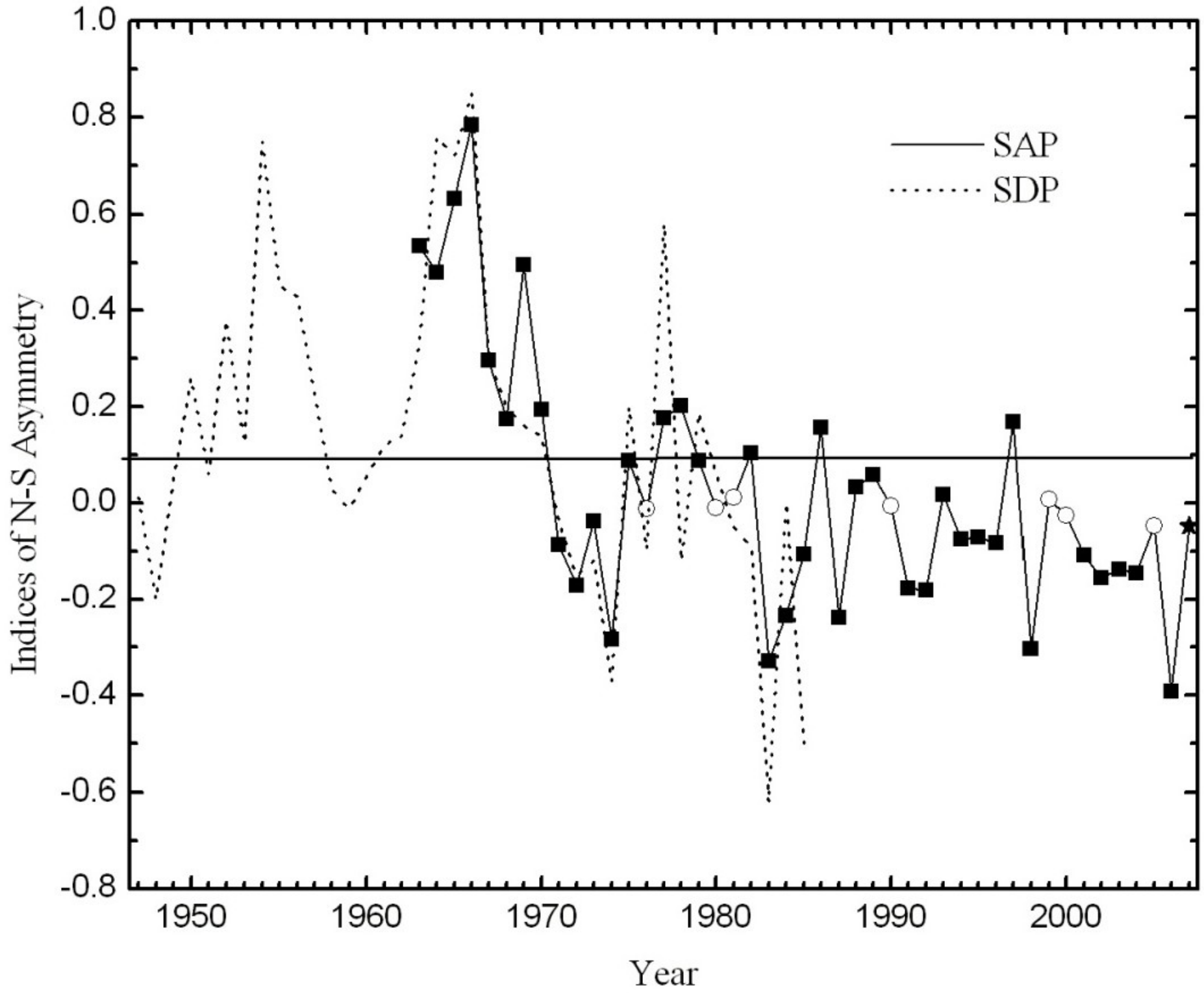

**Figure 4.** Plot of the N-S asymmetry indices of SAP events versus years (1963-2007). Highly significant, significant and insignificant values are marked with black squares (■), black stars (★) and white circles (○) respectively.

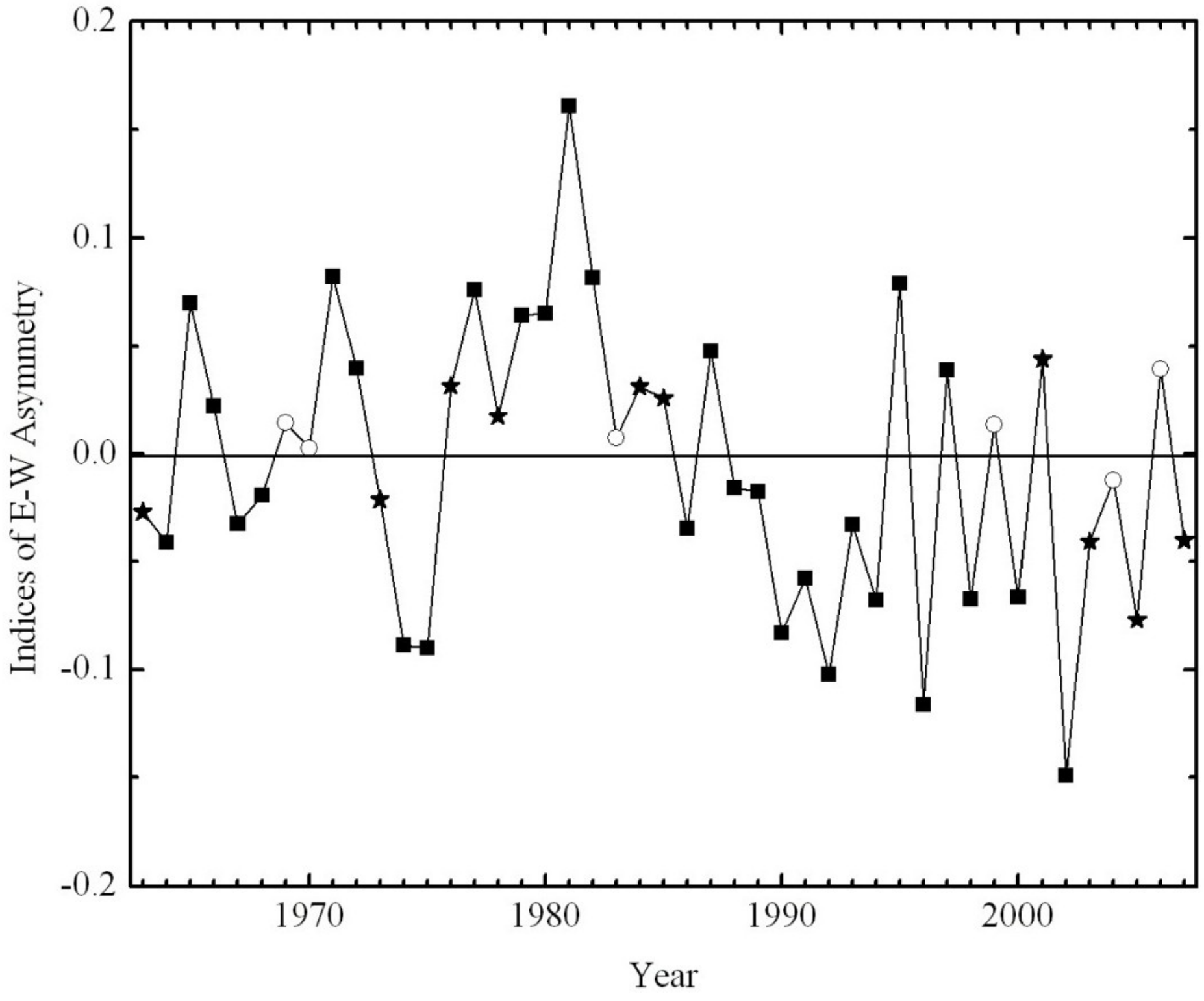

**Figure 5.** Same as Figure 4, but for E-W asymmetry.